\documentclass[
showpacs,preprintnumbers,amsmath,amssymb]{revtex4}
\usepackage{amsmath}
\usepackage{empheq}
\usepackage{graphicx}
\begin{document}

\title{INFLATIONARY COSMOLOGY LEADING TO A SOFT TYPE SINGULARITY}

\author{   I. Brevik$^{1}$\footnote{E-mail:iver.h.brevik@ntnu.no}, V. V. Obukhov$^{2}${\footnote{E-mail:obukhov@tspu.edu.ru}},
   A. V. Timoshkin$^{3}${\footnote{E-mail:timoshkinAV@tspu.edu.ru}}. }

\medskip

\affiliation{$^{1}$Department of Energy and Process Engineering, Norwegian University
of Science and Technology, N-7491 Trondheim, Norway}.

\affiliation{$^{2}$Tomsk State Pedagogical University,   Kievskaja Street 60,  634050 Tomsk, Russia}

\affiliation{$^{3}$Tomsk State Pedagogical University, Kievskaja Street 60, 634050 Tomsk, Russia;
National Research Tomsk State University, 634050 Tomsk, Russia;
Laboratory of Theoretical Cosmology, Tomsk State University of Control Systems and Radio-electronics, Lenin Avenue 40, 634050 Tomsk, Russia}.

 \today

\begin{abstract}
A remarkable property of modern cosmology is that it allows for a special case of symmetry, consisting in the possibility of describing the early-time acceleration (inflation) and the late-time acceleration using the same theoretical framework. In this paper we consider various cosmological models corresponding to a  generalized form for the equation of state for the fluid in  a flat Friedmann -Robertson-Walker universe, emphasizing  cases where the  so-called type IV singular inflation is encountered in the future. This is  a soft (non-crushing) kind of singularity.  Parameter values for an inhomogeneous equation of state leading to singular inflation are obtained. We present models for which there are two type IV singularities, the first corresponding to the end of the inflationary era  and the second to a late time event. We also study the correspondence between the theoretical  slow-roll parameters leading to  type IV singular inflation and the recent results observed by the Planck satellite.

\end{abstract}

 \maketitle
\section{Introduction}

The observed  late-time accelerated expansion of the universe \cite{riess99,perlmutter99} led to the emergence of new cosmological models \cite{nojiri05}. A full description of cosmological evolution ought to include the early-time acceleration just after the inflationary period \cite{linde08,gorbunov11}.  The inflationary epoch is characterized by  total energy as well as  scale factor increasing exponentially \cite{brandenberger10}. This phenomenon can be described in terms of a cosmic fluid  satisfying an inhomogeneous equation of state, and leads to physical properties different from those found for standard matter and radiation.
Such a generalized fluid can  be made use of in the modeling of the inflationary era \cite{bamba15}.  The inclusion of inhomogeneous fluids in cosmology can be looked upon as a way of adopting   modified gravity \cite{nojiri11} - the fluid exists in the gravitational field obeying an unconventional equation of state \cite{capozziello06}.

Various types of finite time singularities can occur both after early-time acceleration, and after late-time acceleration.  It is of interest to study the cosmological evolution of the universe that is associated with the so-called type IV finite time singularities. The finite time cosmological singularities were first classified in Ref.~\cite{nojiri05}. Type IV singularity is the mildest singularity among all the singularities, because  physical quantities such as the scale factor, the effective energy density, and the pressure, remain finite whereas  the higher derivatives of the Hubble rate diverge. These are thus  not crushing type singularities. The universe's evolution continues smoothly after having passed through the  singularity. The  finite time future singularity, more often considered in the literature,  is by contrast a sudden singularity. Singularities of this kind, of type II, have been studied in Refs.~\cite{bamba10,yurov08}.  Recently, the idea of a singular inflationary universe was proposed in  Refs.~\cite{nojiri15,nojiri15a,odintsov15}, with occurrence of a type IV singularity at the end of the inflation.

The purpose of this article is to study the singular inflation for  types II and IV cosmological evolution induced by generalized equation-of-state fluids. We will investigate different general forms of the Hubble rate that can lead to future  singularities. Some concrete examples of evolution, following from specific choices of parameters in the inhomogeneous equation of state, are considered. In some of the models there are two type IV singularities, the first one corresponding to inflation, the second one to a late time event. This kind of symmetry is a characteristic kind of symmetry that has recently emerged in modern cosmology. We calculate the slow-roll parameters, the spectral index, and the tensor-to-scalar ratio.  Singularities of the slow-roll-parameters can be traced back to some sort of dynamic instability in the governing equations with some specific choices of the parameters. Finally we  discuss how the theoretical   spectral indices compare with the Planck observational data.

\section{	Singular inflation models from modified equation of state}

Let us consider a  perfect fluid in standard Einstein-Hilbert gravity and write the  Friedmann-Robertson-Walker (FRW) equation as
\begin{equation}
\rho=\frac{3}{k^2}H^2, \label{1}
\end{equation}
where $\rho$  is the energy density,  $H(t)=\dot{a}(t)/a(t)$ is  the Hubble parameter, $a(t)$  is the scale factor, and $k^2=8\pi G$  with  $G$ being Newton's gravitational constant (a dot  denotes derivative with respect to  cosmic time $t$).

We assume that the FRW metric is flat and has the  form
\begin{equation}
ds^2=-dt^2+a^2
(t)\sum_i dx_i^2. \label{2}
\end{equation}
We assume that our universe can be described by a fluid that obeys a non-linear inhomogeneous equation of state depending on time,
\begin{equation}
p=w(t)\rho+f(\rho)+\Lambda(t), \label{3}
\end{equation}
where  $p$  is the pressure,  $w(t)$  and $\Lambda(t)$   depend on time $t$,  and $f(\rho)$   is an arbitrary function in the general case.

Generally speaking, an effective equation of state of this class is typical in modified gravity (see Refs.~\cite{nojiri07,nojiri03} for  reviews). In the present paper  paper we will investigate the phenomenological equation of state (\ref{3}) for the so-called  types II and IV cosmological evolutions occurring in the inflationary epoch.

Here we recall the classification of finite time cosmological singularities \cite{nojiri05}:
\begin{itemize}
\item 	Type II ("sudden"): when $t \rightarrow t_s$, the scale factor and the effective energy density are finite, that is $a\rightarrow a_s$, $\rho_{\rm eff} \rightarrow \rho_s$,  , but the effective pressure diverges, i.e. $|p_{\rm eff}|\rightarrow \infty$.
\item	Type IV: when  $t\rightarrow t_s$,  the scale factor, the effective energy density and the effective pressure are finite, that is $a\rightarrow a_s, \rho_{\rm eff}\rightarrow \rho_s, |p_{\rm eff}|\rightarrow p_s$, but the higher derivatives of the Hubble rate diverge.
\end{itemize}
Now  consider a general form of the Hubble parameter that can produce  type II and type IV singular cosmological evolutions \cite{nojiri15b}
\begin{equation}
H(t)=f_1(t)+f_2(t)(t_s-t)^\alpha, \label{4}
\end{equation}
where the functions  $f_1(t)$ and $f_2(t)$ are smooth and differentiable.  If the values of the parameter $\alpha$  are restricted  to $0<\alpha <1$, the cosmological solution  develops a type II singularity, while when  $\alpha >1$ it develops a  type IV singularity. We will take the parameter $\alpha$ to have the following form:
\begin{equation}
\alpha =\frac{m}{2n+1}, \label{5}
\end{equation}
where  $n,m \in N$.
 	In general case it is difficult to find a corresponding form equation of state for the fluid when $f_1(t)$ and $f_2(t)$ are assigned arbitrary analytic forms.  For simplicity we will here assume that the function $f(\rho)=0$ and moreover assume  the parameters $w(t)$ and $\Lambda(t)$   to be linearly dependent on time \cite{brevik07}, i.e.
 \begin{equation}
 w(t)=at+b, \quad \Lambda(t)=ct+d, \label{6}
 \end{equation}
(this is a very reasonable choice in view of the latest observational proposals for a an  effective equation-of-state parameter). Here $a,b,c,d$   are arbitrary constants.

Let us begin with a simple example, choosing the functions as $f_1(t)=f_0$  and  $f_2(t)=g_0$,  where $f_0$ and $g_0$  are  arbitrary positive dimensional constants. At first we  investigate the case $\alpha=2$ in Eq.~(\ref{4}), which then describes  type IV evolution.

The  energy conservation law is
\begin{equation}
\dot{\rho}+3H(\rho+p)=0, \label{7}
\end{equation}
and with use of Eqs.~(\ref{1}), (\ref{3}), (\ref{4}) and (\ref{6}) we then get
\begin{equation}
4(t-t_s)g_0+3(at+b+1)\left[ f_0+g_0(t-t_s)^2\right]^2+k^2(ct+d)=0. \label{8}
\end{equation}
From this equation  the constants  $a,b,c,d$ can be determined. By substituting Eq.~(\ref{6}) in the effective pressure equation (\ref{3}) we obtain the equation of state
\begin{equation}
p=-\rho-\frac{2g_0}{k^2}t+\frac{g_0t_s}{k^2}(2+3g_0t_s^3). \label{9}
\end{equation}
This is thus  one specific example of an inhomogeneous phenomenological equation of state belonging to the general class of Eq.~(\ref{3}).

Now we will study a singular inflationary model, leading to a type II cosmological evolution, by setting  $\alpha=1/2$ in Eq.~(\ref{4}). In this case the energy conservation law takes the form
\begin{equation}
g_0\sqrt{t_s-t}-3(at+b+1)\left[ f_0+g_0\sqrt{t_s-t}\right]^2-k^2(ct+d)=0. \label{10}
\end{equation}
 If we take the constants in Eq.~(\ref{10}) to have the following values:
 \begin{equation}
 \left\{ \begin{array}{llll}
 \alpha=0, \\
 b=-r_0+\frac{k^2}{3g_0^2}c,  \\
 d=\frac{1}{k^2}\left\{ \frac{g_0}{\sqrt{t_s}}-3\left[h_0+b\left( h_0+2f_0g_0\sqrt{t_s}\right)\right]\right\},  \\
 c \in R_+,
 \end{array}
 \right. \label{11}
 \end{equation}
 then we obtain type II cosmological evolution.

In particular, if $c=0$  the modified equation of state can be written in the following form:
\begin{equation}
p=-r_0\rho+\frac{1}{k^2}\left[ \frac{g_0}{\sqrt{t_s}}-3\left(h_0-r_0(h_0+2f_0g_0\sqrt{t_s})\right)\right], \label{12}
\end{equation}
where $h_0=f_0^2-\frac{f_0g_0}{\sqrt{t_s}} +g_0^2t_s, r_0=1+\frac{1}{6g_0t_s^{3/2}}, $ and $|t| \leq t_s$.

Now we consider the following model \cite{nojiri15b}:
\begin{equation}
H(t)=h_0\left[ \left(\frac{t-t_0}{t_1}\right)^{-2n}+1\right]^{-\frac{\alpha}{2n}}, \label{13}
\end{equation}
where $h_0,t_0,t_1,\alpha$ are constants.  We assume that $h_0>0,n>0$, and $\alpha>0$.  In this case a type IV singularity occurs, since the Hubble parameter behaves as $H(t) \approx h_0\left(\frac{t-t_0}{t_1}\right)^\alpha$.

We will now investigate the particular case when  $\alpha=2n, n=1/2$.  The corresponding energy conservation law takes the form
\begin{equation}
\frac{2h_0t_1}{k^2(t+t_1-t_0)^2}+\frac{3h_0^2}{k^2}(at+b+1)\frac{(t-t_0)^2}{(t+t_1-t_0)^2}+ct+d=0. \label{14}
\end{equation}
The following parameters  satisfy this equation:
\begin{equation}
\left\{ \begin{array}{llll}
\alpha=0, \\
c=2[\tau_1+\tau_2(b+1)], \\
d=\frac{1}{2}(t_0-t_1)c, \\
b \in R,
\end{array}
\right. \label{15}
\end{equation}
where $\tau_1=\frac{2h_0t_1}{k^2(t_1-t_0)^3}, \tau_2=\frac{3h_0^2}{k^2(t_1-t_0)}$,
and $\left| \frac{t}{t_1-t_0}\right| <1$.

If  $b=-1$ we obtain the following approximate equation of state:
\begin{equation}
p=-\rho +2\tau_1\left( t+\frac{t_0-t_1}{2}\right). \label{16}
\end{equation}
Next, let us consider the case in which the Hubble parameter is given by \cite{nojiri15b}
\begin{equation}
H(t)=f_0(t-t_1)^\alpha+c_0(t-t_2)^\beta, \label{17}
\end{equation}
where  $f_0$ and $c_0$ are constant positive parameters,  and  $\alpha,\beta >1$.  In this model
the cosmological evolution has two type IV singularities, at  $t=t_1$  and  $t=t_2$.  We suppose that  $t_1$ is at the end of the inflationary era,  and $t_2$   is a late time. In the general case it is difficult  to obtain the exact  equation of state for the Hubble parameter \cite{nojiri15b}. Therefore we will consider the behavior of the cosmological system in the vicinity of the early-time, and the late-time, type IV singularity.

So, in the first case at $t \approx t_1$   the equation of motion takes the form
\begin{equation}
-2c_0\beta (t-t_2)^{-1+\beta}=3c_0^2(at+b+1)(t-t_2)^{2\beta}+k^2(ct+d). \label{18}
\end{equation}
For example, with $\beta=2$   we obtain
\begin{equation}
\left\{ \begin{array}{llll}
a=0, \\
b=-1,  \\
c=-4c_0/k^2,  \\
d=-ct_2,
\end{array}
\right. \label{19}
\end{equation}
and  the equation of state takes the approximate form
\begin{equation}
p=-\rho-\frac{4c_0}{k^2}(t-t_2). \label{20}
\end{equation}
Thus it appears that the late-time singularity controls the early-time inhomogeneous equation of state.

Analogously, near the late-time type IV singularity at $t\approx t_2$, if we let $\beta \rightarrow \alpha$ in Eq.~(\ref{18}) and consider the particular case $\alpha =2$ , we obtain the following equation of state, symmetric to Eq.~(\ref{20}),
\begin{equation}
p=-\rho-\frac{4c_0}{k^2}(t-t_1). \label{21}
\end{equation}
Thus,  the equation of state  near the late-time singularity is solely controlled by the early-time type IV singularity.

Let us now go over to another interesting model which contains a unified description of inflation at early time with a late-time acceleration of the universe. At late time a  type IV singularity occurs. The Hubble parameter in this model is \cite{nojiri15b}
\begin{equation}
H(t)=\frac{f_1}{\sqrt{t^2+t_0^2}}+\frac{f_2t^2(-t+t_1)^\alpha}{t^4+t_0^4}+f_3(-t+t_2)^\beta. \label{22}
\end{equation}
The parameters  $\alpha,\beta,t_0,f_1,f_2,f_3$  are positive constants, so that we have $H(t)>0$.  The time $t_1$ is associated with early time, while  $t_2$ is associated with late time. If $\alpha >1$ and $\beta>1$ a type IV singularity occurs at both early time and at late time (cf. the previous example).

Let us find, as in the previous case, an analytic approximation for the equation of state near the singularities.
At first, we consider the case when the physical system lies near the early-time singularity, that is $t \approx t_1$.  We have then the following form for  the energy conservation equation:
\begin{align}
& \frac{2f_1t}{(t^2+t_0^2)^{3/2}}-\frac{3f_1^2}{t^2+t_0^2}-
\frac{6f_1f_3(-t+t_2)^\beta}{(t^2+t_0^2)^{1/2}}-f_3(-t+t_2)^\beta\left[3f_3(-t+t_2)^\beta-\frac{2\beta}{t^2+t_0^2}\right] \notag \\
& = (at+b)\left[\frac{3f_1^2}{t^2+t_0^2}+\frac{6f_1f_3(-t+t_2)^\beta}{(t^2+t_0^2)^{1/2}}+3f_3^2(-t+t_2)^{2\beta}\right]+k^2(ct+d). \label{23}
\end{align}
Let us consider the case $\beta =2$, and let us put $f_1=0$. The equation of state then becomes approximatively
\begin{equation}
p=-\rho-\frac{4f_3}{k^2}(t-t_2). \label{24}
\end{equation}
Thus the late-time type IV singularity controls again, as in the example above, the behavior of the cosmological dark fluid near the early-time type IV singularity. If we simplify the expression (\ref{24}) by supposing $t\approx t_2$, then the equation of state reads $p\approx -\rho$. This means that the early-time  evolution is close to the  de Sitter acceleration.

If we put in Eq.~(\ref{24}) the parameter $f_3=0$, we get a different form for the the equation of state:
\begin{equation}
p=\left( -1+\frac{t}{f_1t_0}\right)\rho-\frac{f_1}{k^2t_0^3}t, \label{25}
\end{equation}
where $|t| \leq t_0$.

Considering conversely the behavior of the cosmological system at late time and near the future type IV singularity, that is $t\approx t_2$, we find the corresponding equation of state to have the form
\begin{equation}
\frac{2f_1t}{(t^2+t_0^2)^{3/2}}=(at+b+1)\frac{3f_1^2}{t^2+t_0^2}+k^2(ct+d). \label{26}
\end{equation}
We choose here $\alpha=2$, and simplify the energy conservation equation by  taking into account that  $f_0=0.$ The corresponding equation of state becomes
\begin{equation}
p=\left( \frac{t}{3f_1t_0}-1\right) \rho+\frac{f_1}{k^2t_0^3}t, \label{27}
\end{equation}
where $|t| \leq t_0$.

Thus, we have found analytic approximations for the equation of state near the singularities.

\section{Confronting singular inflationary models with observational data}

In the previous section we  considered various representations of singular inflation cosmological models in the presence of a type IV singularity via the modified equation of state. The presence of a type IV singularity in the inflationary stage of the evolution of the universe can have an influence upon the  observed indices. At first, we will calculate the slow-roll parameters as  functions of the cosmic time in the general case. Then, we will consider an illustrative example where the Hubble has the form (4), and will compare some of the results  with the recent Planck observational data \cite{ade14}. The analysis will depend strongly on the concrete model, which is related to the type IV singularity. A detailed analysis of the slow-roll parameters for a perfect fluid,  described with a phenomenological equation of state, is given in  Ref.~\cite{bamba14}.

Let us calculate the slow-roll parameters $\varepsilon$ and $\eta$. We obtain
\[
\varepsilon = -\frac{H^2}{4\dot{H}}\left( \frac{6\dot{H}}{H^2}+
\frac{\ddot{H}}{H^3}\right)^2\left(3+\frac{\dot{H}}{H^2}\right)^{-2},   \]
\begin{equation}
 \eta =-\frac{1}{2}\left( 3+\frac{\dot{H}}{H^2}\right)^{-1}\left(  \frac{6\dot{H}}{H^2}+\frac{{\dot{H}}^2}{2H^4}-\frac{\ddot{H}}{H^3}-\frac{{\dot{H}}^4}{2H^4}+
\frac{{\dot{H}}^2\ddot{H}}{H^5}-\frac{{\ddot{H}}^2}{2H^2}+\frac{3{\ddot{H}}}{H\dot{H}}+\frac{\dddot{H}}{H^2\dot{H}}\right),\label{28}
\end{equation}
\[ \xi^2 = \frac{1}{4}\left( \frac{6\dot{H}}{H^2}+
\frac{\ddot{H}}{H^3}\right)\left(3+\frac{\dot{H}}{H^2}\right)^{-1}
\left( \frac{9\ddot{H}}{H\dot{H}}+\frac{3\dddot{H}}{H^2}+\frac{2\dddot{H}}{H^2\dot{H}}+\frac{4{\ddot{H}}^2}{H^2{\dot{H}}^2}
-\frac{\ddot{H}\dddot{H}}{H{\dot{H}}^3}-\frac{3{\ddot{H}}^2}{{\dot{H}}^3}+
\frac{{\ddot{H}}^3}{H{\dot{H}}^4}+
\frac{\ddddot{H}}{H{\dot{H}}^2}\right).  \]
Now we assume, as an example, that the Hubble parameter has the form of Eq.~(\ref{4}), and we consider the simple case where $H(t)=g_0(t-t_s)^2$. We then obtain
\begin{align}
 &\varepsilon =\frac{g_0}{2}\Delta t_s\left( \frac{6\Delta t_s+1}{3g_0\Delta t_s^3+2}\right)^2, \notag \\
&\eta = \frac{15g_0+12g_0^2\Delta t_s^{-1}-2(t_s^2+8)\Delta t_s^{-5}}{2g_0(3g_0\Delta t_s^3+2)}, \notag \\
&\xi^2 =\frac{5(6\Delta t_s+1)(3g_0\Delta t_s^3+1)}{4g_0^2\Delta t_s^7(3g_0\Delta t_s^3+2)}, \label{29}
\end{align}
with $\Delta t_s=t-t_s$. From these expressions we see that there are singularities present for $t=t_s$, which corresponds to type IV singularity. The slow-roll singularities indicate an instability of the dynamical system that corresponds to this cosmological model.

 Finally, we will choose a suitable phenomenological function in the equation of state  and compare the results of our analysis with recent Planck \cite{ade14}.  We will consider the following observational indices: the spectral index of primordial curvature perturbations $n_s$, the scalar-to-tensor ratio  $r$, and the running  spectral index $a_s$. For reasons of comparison we  will consider a simple expositional model, which leads to a type IV singularity. We take the phenomenological equation of state in the form
 \begin{equation}
 p=-\rho+f(\rho), \label{30}
 \end{equation}
 where  $f(\rho)=A\rho^\alpha$, $A$ and $\alpha$ being  positive constants. The effective energy density as a function of the $e$ -folding  $N$ is equal to \cite{nojiri15b}
 \begin{equation}
 \rho=[3(1-\alpha)A]^{\frac{1}{1-\alpha}}N^{\frac{1}{1-\alpha}}. \label{31}
 \end{equation}
As was demonstrated in Ref.~\cite{nojiri15b}, a type IV singular evolution takes place when the parameter $\alpha$  is restricted to the interval $0<\alpha <1/2$. If the fraction
\begin{equation}
\frac{f(\rho)}{\rho} \leq 1, \label{32}
\end{equation}
the observational indices can be approximated by the following expressions \cite{bamba14}
\begin{equation}
n_s \approx 1-6\frac{f(\rho)}{\rho(N)}, \quad r\approx 24\frac{f(\rho)}{\rho(N)}, \quad a_s\approx -9\left(\frac{f(\rho)}{\rho(N)}\right)^2. \label{33}
\end{equation}
Taking into account Eqs.~(\ref{31}), (\ref{32}) and (\ref{33}) we obtain the approximate expressions for the observational indices:
\begin{equation}
n_s\approx 1-\frac{2}{N(1-\alpha)}, \quad r\approx \frac{8}{N(1-\alpha)}, \quad a_s\approx -\frac{1}{N^2(1-\alpha)^2}. \label{34}
\end{equation}
If we use the values $(N, \alpha)=(70, 1/10)$, the observational indices become
\begin{equation}
n_s\approx 0.9683, \quad r\approx 0.127, \quad a_s\approx -0.00025. \label{36}
\end{equation}
The Planck data \cite{ade14} give the following observational indices:
\begin{equation}
n_s=0.9644\pm 0.0049, \quad r<0.10, \quad a_s=-0.0057\pm 0.0071. \label{36}
\end{equation}
 We see that there is good  agreement with the spectral index of primordial curvature perturbations. However, the scalar-to-tensor ratio is not so well predicted, and a more definite  disagreement is seen to occur for   the running spectral index. These mismatches  can be removed, if we consider more sophisticated   models than the  simple  one used here.

 \section{Conclusion}

 We have considered various phenomenological equations of state  leading  to a type IV cosmological evolution. Such  models are typical when considering  acceleration of the universe, because the equation-of state-parameter $w$ is close to $-1$.  This type of  singularity is the mildest one among  the singular solutions, and   is of a non-crushing nature. We have shown how some models contain two singularities, one corresponding to early time (inflation), and one to late time, thus characterizing an important symmetry property contained in modern cosmology.
  We have discussed in an   example how the theoretical  spectral indices compare with the Planck observation data. It is noteworthy  how also  the slow-roll parameters can become singular in some cases, thus influencing the Hubble parameter and consequently  influencing the evolution of the universe. The physical meaning of this is  that instabilities can occur in the dynamical system.

 An interesting point is that  type IV singularity may  provide a very natural scenario for graceful exit, as proposed recently in Ref.~\cite{odintsov15a}.

Our theory can be extended to the case of a type IV singular evolutionary theory where the coupling with matter is included.

\section*{Acknowledgement}
              This work was supported by a grant from the Russian Ministry of Education and Science, project TSPU-139 (A.V.T.).
              
              \bigskip
              \noindent {\bf Note added.} The following reference should have been included: Barrow, J. D. and Graham, A. A. H. Singular inflation, Phys. Rev. D {\bf 91}, 083513 (2015). It shows how a physicallly motivated equation of state (scalar field with a power-law potential) leads to a sudden finite-time singularity.

\end{document}